# Modelling vitamin D food fortification among Aboriginal and Torres Strait Islander peoples in Australia

Belinda Neo [1], Noel Nannup [2], Dale Tilbrook [3], Eleanor Dunlop [4,5], John Jacky [2], Carol Michie [2], Cindy Prior [2], Brad Farrant [2], Carrington C.J. Shepherd [1,2,6], and Lucinda J. Black [4,5,*]

[1] Curtin Medical School, Curtin University, Bentley, Western Australia, Australia
[2] Telethon Kids Institute, The University of Western Australia, Nedlands, Western Australia, Australia
[3] Maalingup Aboriginal Gallery, Caversham, Western Australia, Australia
[4] Curtin School of Population Health, Curtin University, Bentley, Western Australia, Australia
[5] Institute for Physical Activity and Nutrition (IPAN), School of Exercise and Nutrition Sciences, Deakin University, Geelong, Victoria, Australia
[6] Ngangk Yira Institute, Murdoch University, Murdoch, Western Australia, Australia
* Correspondence: Lucinda J. Black Lucinda.black@deakin.edu.au

**Abstract**

**Background**: Low vitamin D intake and high prevalence of vitamin D deficiency (serum 25-hydroxyvitamin D concentration < 50 nmol/L) among Aboriginal and Torres Strait Islander peoples highlight a need for public health strategies to improve vitamin D status. As few foods contain naturally occurring vitamin D, fortification strategies may be needed to improve vitamin D intake and status among Aboriginal and Torres Strait Islander peoples.

**Objective:** We aimed to model vitamin D food fortification scenarios among Aboriginal and Torres Strait Islander peoples.

**Methods:** We used nationally representative food consumption data (n=4,109) and vitamin D food composition data to model four food fortification scenarios. The modelling for Scenario 1 included foods and maximum vitamin D concentrations permitted for fortification in Australia: i) dairy products and alternatives, ii) butter/margarine/oil spreads, iii) formulated beverages, and iv) selected ready-to-eat breakfast cereals. The modelling for Scenarios 2a-c included some vitamin D concentrations higher than permitted in Australia; Scenario 2c included bread, which is not permitted for vitamin D fortification in Australia. Scenario 2a: i) dairy products and alternatives, ii) butter/margarine/oil spreads, iii) formulated beverages. Scenario 2b: as per Scenario 2a plus selected ready-to-eat breakfast cereals. Scenario 2c: as per Scenario 2b plus bread.

**Results:** Vitamin D fortification of a range of staple foods could potentially increase vitamin D intake among Aboriginal and Torres Strait Islander peoples by ~ 3-6 µg/day. Scenario 2c showed the highest potential median vitamin D intake increase to ~ 8 µg/day. Across all modelled scenarios, none of the participants had vitamin D intake above the Australian upper level of intake of 80 µg/day.

**Conclusions:** Vitamin D fortification of commonly consumed staple foods could safely increase vitamin D intake among Aboriginal and Torres Strait Islander peoples, even considering the current restrictions on the foods permitted for fortification in Australia.

**Introduction**

Vitamin D deficiency (serum 25-hydroxyvitamin D [25(OH)D] concentration < 50 nmol/L) is a prevalent public health issue in Australia, affecting 27% of all Aboriginal and Torres Strait Islander peoples aged ≥ 18 years and 39% of those living in remote areas (1). Vitamin D deficiency may result in adverse health outcomes in musculoskeletal health and other non-communicable diseases (2-4). Vitamin D deficiency can be attributed to a lack of ultraviolet-B radiation via sun exposure and low dietary vitamin D intake (5). In Australia, sun-protective behaviours are influenced by important public health initiatives designed to reduce the high incidence of melanoma (6). Mean vitamin D intake quantified among the general Australian population, which included a few randomly selected Aboriginal and Torres Strait Islander peoples, was reported to be low at <3.5 µg/day (5). The estimated average requirement of vitamin D intake recommended by the Institute of Medicine is 10 µg/day, assuming minimal sunlight exposure (7). Given the prevalence of vitamin D deficiency and low dietary intake, there is a need for public health strategies to improve vitamin D intake and status among Aboriginal and Torres Strait Islander peoples.

Food fortification can increase the dietary supply of nutrients without requiring a change in consumption behaviours (8). As there are very few naturally rich food sources of vitamin D, adding vitamin D to staple foods may be an appropriate strategy to increase intake at the population level (9, 10). For Aboriginal and Torres Strait Islander peoples living in areas with limited access to traditional foods high in vitamin D (e.g., oily fish), fortifying staple foods would increase the availability of food sources of vitamin D.

In Europe, the US, and Canada, a range of food vehicles are fortified with vitamin D, including dairy products (e.g., milk, yoghurt, butter, and cheese); plant-based fluid milk alternatives (e.g., soy, oat, and almond); bread; margarine; orange juice; and ready-to-eat breakfast cereals (RTEBCs) (11-14). In Finland, following the voluntary but systematic routine addition of vitamin D to fat spreads, dairy products and alternatives (e.g., soy or cereal-based drinks and soy yoghurt), juice/nectars, and water/water-based beverages, the prevalence of vitamin D deficiency reduced from 56% in 2000 to 9% in 2011 (12).

In Australia, vitamin D fortification is mandated for edible oil spreads, such as margarine, at a minimum concentration of 5.5 µg/100 g (15). Other food products, such as milk and yoghurt, are permitted for voluntary fortification with vitamin D (16), but this practice does not occur routinely. Recently, our team modelled a scenario in which the maximum permitted amount of vitamin D (0.8 µg/100 mL) was added to all fluid milk products and plant-based alternatives in Australia (17). The modelling showed an increase in vitamin D intake of ~ 2 µg/day across the Australian population but demonstrated that a more comprehensive vitamin D food fortification strategy across a wider range of food products would be needed for the majority of Australians to achieve an intake of 10 µg/day. Other studies have highlighted that adding vitamin D to a single product or a few staple food products may limit the reach of such interventions across the population distribution (10, 18). We aimed to model vitamin D food fortification scenarios using a range of staple foods in the Aboriginal and Torres Strait Islander population in Australia.

**Methods**

**Study population**

The 2012-2013 National Aboriginal and Torres Strait Islander Nutrition and Physical Activity Survey (NATSINPAS) provided the most recent food consumption data for Aboriginal and Torres

Strait Islander peoples. A detailed description of the methods is available elsewhere (19, 20). In brief, the 2012-2013 NATSINPAS was conducted between August 2012 and July 2013 among Aboriginal and Torres Strait Islander peoples living in remote and non-remote communities. In each dwelling, an adult household member (aged >18 years) was interviewed, and a representative adult household member (child proxy) was interviewed for children aged ≥ 2 years. The first interview was conducted face-to-face (n=4,109), and a second telephone interview (n=771) was conducted eight days after the first interview for those living in non-remote areas. The interview data were recorded electronically using a Computer Assisted Person Interview instrument system to ensure the data were captured accurately and securely. There were 4,088 participants (99.5%) in the first interview, and data for the remaining 21 participants (0.5%) were imputed. As the second interview had a low response rate of 18.8% (n=771) and was only conducted in those living in non-remote areas, we did not use those data in our analysis.

**Food consumption data**

A 24-hour food recall was used to collect dietary intake over the previous 24 hours in participants aged ≥ 2 years. Children aged 6-14 years were encouraged to participate in the 24-hour food recall, and children aged 15-17 years were interviewed with parental or guardian consent. Detailed methods are reported elsewhere (19, 20). In brief, the Automated Multiple-Pass Method (AMPM) was used to collect detailed information about dietary intake through systematic question prompts. The AMPM was developed by the United States of Agriculture and adapted for use in Australia by Food Standards Australia New Zealand. The Australian Health Survey food model booklet (21) and bush tucker prompt cards (22) were used to aid participants in describing the quantities of food consumed. Food consumption data collected were coded using the unique 8-digit food code assigned to each individual food entry in the 2011-2013 AUSNUT (AUStralian Food and NUTrient) Database, which contains nutrient composition data for 5,740 foods recorded in the NATSINPAS (23).

**Vitamin D composition data**

Our team previously published analytical vitamin D food composition data for four D vitamers (vitamin $D_3$, 25(OH)$D_3$, vitamin $D_2$, and 25(OH)$D_2$) in retail food products and game products (24, 25). Retail food products were procured across three cities: Sydney, Melbourne, and Perth. Game products such as camel, farmed crocodile, farmed emu, and wild kangaroo meat were purchased commercially from Yarra Valley Game Meats located in Victoria, and emu eggs and oil were purchased from emu farms located in Western Australia, Victoria, and New South Wales. Samples were analysed using a validated liquid chromatography with triple quadrupole mass spectrometry method at the National Measurement Institute of Australia, a laboratory accredited by the National Association of Testing Authorities for measuring vitamin D in food (ISO17025:2017).

We previously mapped analytical vitamin D concentrations to the 5,740 food entries in the 2011-2013 AUSNUT database, following methods used for the Australian Total Diet Studies (26). Foods were either mapped to single food items (e.g., milk) or recipe foods (containing single food items as ingredients, e.g., custard). A conversion factor was used to adjust the analytical vitamin D values for foods that may undergo food manufacturing or preparation processes, such as the reconstitution of infant formula.

**Food vehicles for fortification**

For our models, we used foods permitted for vitamin D fortification in Australia (dairy products and alternatives, butter/margarine/oil spreads, formulated beverages, and selected RTEBCs) according to the Australia New Zealand Food Standards Code (16), along with bread, which is not permitted for vitamin D fortification in Australia. Bread, dairy milk, breakfast cereals, and cheese are consumed by 70%, 69%, 34%, and 27% of Aboriginal and Torres Strait Islander peoples, respectively (27).

RTEBCs are only permitted for vitamin D fortification in Australia if they meet the Nutrient Profiling Scoring Criterion (NPSC) (28). The NPSC provides a score calculated based on energy, saturated fat, sodium and sugar content, the percentage of fruit and vegetables, and, for some foods, protein and dietary fiber. We used the nutrient information based on 100 g of each RTEBC in the 2011-2013 AUSNUT Food Recipe File to determine whether they were NPSC-compliant. RTEBCs that met the NPSC were included in the modelling scenarios. According to the Australia New Zealand Food Standards Code, the maximum claim for vitamin D concentration is 2.5 μg per normal serving for RTEBC (16). We considered a normal serving to be 30 g, the recommended standard serving for wheat cereal flakes in the Australian Dietary Guidelines (29).

**Food fortification scenarios**

We modelled four food fortification scenarios **(Table 1)**. For each scenario, a new vitamin D food composition dataset was created with adjusted vitamin D concentrations. The AUSNUT Food and Dietary Supplement Classification System (30) was used to identify single food items (e.g., milk) that required adjusted concentrations. For recipe foods, a disaggregated 2011-2013 AUSNUT Food Recipe File (31) was used to identify single food items within a recipe, and the vitamin D concentrations for the recipe were adjusted according to the proportion of the vitamin D-containing ingredient/s within the recipe. Vitamin D concentrations were not modified in homemade foods if the recipe was not provided in the AUSNUT Food Recipe File.

The modelling for Scenario 1 included foods and maximum vitamin D concentrations permitted for fortification in Australia: i) dairy products and alternatives, ii) butter/margarine/oil spreads, iii) formulated beverages, and iv) NPSC-compliant RTEBCs. The modelling for Scenarios 2a-c included some vitamin D concentrations higher than permitted in Australia; Scenario 2c included bread, which is not permitted for vitamin D fortification in Australia. Scenario 2a: i) dairy products and alternatives, ii) butter/margarine/oil spreads, iii) formulated beverages. Scenario 2b: as per Scenario 2a plus NPSC-compliant RTEBCs. Scenario 2c: as per Scenario 2b plus bread.

**Quantifying vitamin D intakes for baseline and fortification scenarios**

Absolute median vitamin D intakes were quantified using day one of the 24-hour food recall data from the 2012-2013 NATSINPAS. Baseline vitamin D intakes and food fortification models were quantified using the amount of individual food consumed by each respondent (g/day) multiplied by the vitamin D content (μg/g) of the food. A person weighting variable developed by the Australian Bureau of Statistics for the NATSINPAS (20) was applied. Statistical analysis was conducted using Stata version 17 (StataCorp) (32).

**Results**

Across all sex and age groups, median vitamin D intake increased by 3.6 μg/day, 3.2 μg/day, 4.2 μg/day, and 5.7 μg/day under Scenarios 1, 2a, 2b, 2c, respectively **(Table 2)**. The greatest increase in median vitamin D intake was observed under Scenario 2c, which included the greatest range of foods.

Children aged 2-3 years had the greatest median vitamin D intake increase across all the scenarios and almost an eight-fold increase in median vitamin D intake when Scenario 2c was modelled. Across sex and age groups, older female adults aged ≥ 71 years had the lowest median baseline vitamin D intake and the smallest increase in median vitamin D intakes across all the fortification scenarios. Male children aged 2-3 years had the greatest increase in median vitamin D intake across all the scenarios, increasing by 6.3 µg/day in Scenario 1, 5.2 µg/day in Scenario 2a, 7.2 µg/day in Scenario 2b, and 8.9 µg/day in Scenario 2c. Across all the modelled scenarios, none of the participants had a vitamin D intake above the Australian upper level of intake of 80 µg/day (33).

**Discussion**

Vitamin D fortification of a range of staple foods among Aboriginal and Torres Strait Islander peoples could potentially increase vitamin D intakes by ~ 3-6 µg/day from a baseline of 2 µg/day. If all foods permitted for vitamin D fortification in Australia were fortified with the maximum permitted amount of vitamin D (Scenario 1), the median vitamin D intake would potentially increase from baseline to 5.6 µg/day. The highest increase in vitamin D intake was observed with Scenario 2c, which included the broadest range of food. Excessive vitamin D intake may result in hypercalcemia and hypercalciuria (34); however, none of the participants had a vitamin D intake that exceeded the Australian upper level of intake under any modelled scenarios.

Our findings suggest that food fortification could be suitable as a public health strategy to increase vitamin D intake and ultimately improve vitamin D status across the Aboriginal and Torres Strait Islander population. We previously conducted a meta-analysis, systematic review, and dose-response analyses, which showed a non-linear increase in serum 25(OH)D concentration when vitamin D fortified foods were consumed (35). A median increase of ~ 3 µg/day of vitamin D intake from fortified foods may result in an increase in mean serum 25(OH)D concentration of 10.1 (95% confidence interval (95% CI): 4.1-16.0) nmol/L in children and 6.8 (95% CI: 5.2-8.3) nmol/L in adults. Consequently, if all foods permitted for fortification were fortified with the maximum permitted vitamin D concentration in Australia, the mean serum 25(OH)D concentration may increase between 10.1 - 16.7 nmol/L for children and 6.8 - 8.9 nmol/L for adults.

Maintaining adequate levels of vitamin D status can reduce the risk of falls and fractures (2, 3). In Australia, about 8% (> 9,000 events/year) of hospitalisation for falls and fractures in older adults aged ≥ 65 years are due to vitamin D deficiency (36). Falls and fractures are common among Aboriginal and Torres Strait Islander peoples, with about 7,500 hospitalisations from falls documented in 2021-2022 (37). Cost-effective savings have been projected in other countries by reducing the risk of osteoporotic fractures with vitamin D food fortification (38, 39). For example, vitamin D and calcium food fortification could save the German health and social care systems approximately 315 million Euro annually by preventing osteoporotic fractures in older female adults aged ≥ 65 years (39). Vitamin D food fortification could have similar cost-saving benefits as a public health strategy in Australia.

We modelled the fortification of a range of staple foods commonly consumed among Aboriginal and Torres Strait Islander peoples. By expanding the selection of permitted food vehicles to include bread, one of the most commonly consumed foods among Aboriginal and Torres Strait Islander peoples (27), median vitamin D intakes increased about four-fold from baseline intakes. We projected the greatest increase in median vitamin D intakes when all foods permitted for fortification in Australia and bread were included in our fortification model, demonstrating that fortifying a range of commonly consumed staple foods could effectively increase vitamin D intakes by catering to diverse food preferences. However, expanding the range of foods permitted for vitamin D fortification to include bread would require a change to the Australia New Zealand Food Standards Code.

Our modelling revealed that the potential increase in vitamin D intake was not equal across all sex and age groups. Despite modelling vitamin D fortification in concentrations above those permitted, and including foods not permitted for fortification in Australia (namely, bread), the median vitamin D intake among older Aboriginal and Torres Strait Islander female adults aged ≥ 71 years remained low at 4.2 µg/day. Older female adults deficient in vitamin D have a greater risk of falls and fractures and adverse musculoskeletal health outcomes (40), suggesting a need for other complementary public health strategies to increase vitamin D intake within this population group. Vitamin D supplementation could be a suitable public health strategy and has shown to be useful in achieving adequate vitamin D intakes alongside the intake of traditional foods in northern Norway, which includes multi-ethnic populations such as the Sami people (41). However, as vitamin D supplement intake among Aboriginal and Torres Strait Islander peoples was reported as low (< 1%, n = 38) in the NATSINPAS, messaging to promote supplementation use among specific groups of Aboriginal and Torres Strait Islander peoples and the safety of implementation would need to be explored.

While our modelling demonstrated that food fortification could potentially increase vitamin D intake among Aboriginal and Torres Strait Islander peoples, obtaining vitamin D through traditional foods should still be prioritised. Food and food systems are an integral part of the culture of Aboriginal and Torres Strait Islander peoples and how they connect with Country (42). Before colonisation, around 250 years ago, Aboriginal and Torres Strait Islander peoples consumed highly nutritious traditional foods, protecting them from many current nutrition-related health conditions (42, 43). The scientific literature regarding traditional foods and the nutritional composition of traditional foods is still limited. Yarning with Elders and Aboriginal and Torres Strait Islander peoples who are knowledge holders within the community is essential to learn more about their traditional food practices. The findings will be beneficial in guiding food-based strategies to improve vitamin D intake within Aboriginal and Torres Strait Islander communities.

A strength of our study was the use of analytical vitamin D food composition data containing four D vitamers to quantify potential vitamin D intakes. As day 2 of the 24-hour recall interview was only conducted in non-remote areas and had limited responses (n=771), only single-day food consumption data were used to quantify median vitamin D intakes in the modelling scenarios. We did not include supplement intake in our models as supplement consumption was low (<1%) among participants, indicating that it would not significantly impact our models. While under-reporting is a common issue in dietary surveys and was evident in the 2012-2013 NATSINPAS (44), under-reported foods typically comprise non-nutritive energy-dense foods, such as cakes and confectionary, that are low in vitamin D (45). While the food consumption data are from 2012-2013, it represents the most recently available food consumption data for this population. We did not model scenarios in the general Australian population. However, vitamin D deficiency affects 20% of Australian adults aged ≥ 25 years (46), and vitamin D intakes are also low (<3.5 μg/day) (5); hence, food fortification is likely to benefit the general population as well as the Aboriginal and Torres Strait Islander population.

Our findings demonstrate that fortifying a range of commonly consumed staple foods with vitamin D has the potential to increase vitamin D intake among Aboriginal and Torres Strait Islander peoples. For maximum benefit, a revision of the Australia New Zealand Food Standards Code would be required to permit higher levels of vitamin D concentration for fortification in milk, butter, margarine, and edible oil spreads and to expand the list of foods permitted for vitamin D fortification to include bread. Modelling data from the general Australian population would be needed to demonstrate the impact of any change in policy across the nation.

**Acknowledgments**

We would like to acknowledge, with thanks, Dr. Priscila Machado for her provision of the disaggregated 2011-2013 AUSNUT Food Recipe File for this study.

**Author contributions:** BN, LJB, and ED designed the research; BN conducted research, analyzed data, and wrote the original paper; LJB, ED, DT, NN, JJ, BF, CM, CP and CS reviewed and edited the paper; LJB, ED, NN, DT, JJ and CS provided supervision. LJB had primary responsibility for the final content. All authors read and approved the final manuscript.

**Data Availability*:**

Data described in the manuscript were sourced from publicly available datasets found here: 2012-2013 National Aboriginal and Torres Strait Islander Nutrition and Physical Activity Survey https://www.abs.gov.au/statistics/microdata-tablebuilder/microdatadownload, AUSNUT 2011-2013 https://www.foodstandards.gov.au/science-data/monitoringnutrients/afcd/australian-food-composition-database-download-excel-files#nutrient and vitamin D food composition data https://www.foodstandards.gov.au/science-data/monitoringnutrients/afcd/Data-provided-by-food-companies-and-organisations.

**Funding:** This study was supported by the National Health and Medical Research Council (GNT1184788). BN is supported by a Curtin Strategic Scholarship.

**Author Disclosures:** The authors declare no conflict of interest

**References**

1. Black LJ, Dunlop E, Lucas RM, Pearson G, Farrant B, Shepherd CCJ. Prevalence and predictors of vitamin D deficiency in a nationally representative sample of Australian Aboriginal and Torres Strait Islander adults. Br J Nutr. 2021;126(1):101-9.

2. Lips P, van Schoor NM. The effect of vitamin D on bone and osteoporosis. Best Pract Res Clin Endocrinol Metab. 2011;25(4):585-91.

3. Pludowski P, Holick MF, Pilz S, Wagner CL, Hollis BW, Grant WB, et al. Vitamin D effects on musculoskeletal health, immunity, autoimmunity, cardiovascular disease, cancer, fertility, pregnancy, dementia and mortality—A review of recent evidence. Autoimmun Rev. 2013;12(10):976-89.

4. Liu D, Meng X, Tian Q, Cao W, Fan X, Wu L, et al. Vitamin D and multiple health outcomes: an umbrella review of observational studies, randomized controlled trials, and Mendelian randomization studies. Adv Nutr. 2022;13(4):1044-62.

5. Dunlop E, Boorman JL, Hambridge TL, McNeill J, James AP, Kiely M, et al. Evidence of low vitamin D intakes in the Australian population points to a need for data-driven nutrition policy for improving population vitamin D status. J Hum Nutr Diet. 2022;36(1):203-15.

6. Tabbakh T, Volkov A, Wakefield M, Dobbinson S. Implementation of the SunSmart program and population sun protection behaviour in Melbourne, Australia: results from cross-sectional summer surveys from 1987 to 2017. PLoS Med. 2019;16(10):e1002932.

7. Institute of Medicine. Dietary reference intakes for calcium and vitamin D. Washington, DC: National Academies Press; 2011.

8. World Health Organization, Food and Agriculture Organization of the United Nations. Guidelines on food fortification with micronutrients. Lindsay A, de Benoist B, Dary O, Hurrell R, editors. Geneva, Switzerland: World Health Organization and Food and Agriculture Organization of the United Nations; 2006.

9. Lips P, Cashman KD, Lamberg-Allardt C, Bischoff-Ferrari HA, Obermayer-Pietsch B, Bianchi ML, et al. Current vitamin D status in European and Middle East countries and strategies to prevent vitamin D deficiency: a position statement of the European Calcified Tissue Society. Eur J Endocrinol. 2019;180(4):P23-P54.

10. Kiely M, Black LJ. Dietary strategies to maintain adequacy of circulating 25-Hydroxyvitamin D concentrations. Scand J Clin Lab Invest Suppl. 2012;72(sup243):14-23.

11. Calvo MS, Whiting SJ. Survey of current vitamin D food fortification practices in the United States and Canada. J Steroid Biochem Mol Biol. 2013;136:211-3.

12. Jääskeläinen T, Itkonen ST, Lundqvist A, Erkkola M, Koskela T, Lakkala K, et al. The positive impact of general vitamin D food fortification policy on vitamin D status in a representative adult Finnish population: evidence from an 11-y follow-up based on standardized 25-hydroxyvitamin D data. Am J Clin Nutr. 2017;105(6):1512-20.

13. Vatanparast H, Longworth ZL. How does Canada's new vitamin D fortification policy affect the high prevalence of inadequate intake of the vitamin? Appl Physiol Nutr Metab. 2023;48(11):870-5.

14. Pilz S, März W, Cashman KD, Kiely ME, Whiting SJ, Holick MF, et al. Rationale and plan for vitamin D food fortification: a review and guidance paper. Front Endocrinol (Lausanne). 2018;9:373.

15. Food Standards Australia New Zealand. Australia New Zealand Food Standards Code – Standard 2.4.2 – Edible oil spreads [Internet]. Federal Register of Legislation; 2016 [cited 2024 Feb 25]. Available from: https://www.legislation.gov.au/Details/F2016C00174.

16. Food Standards Australia New Zealand. Australia New Zealand Food Standards Code – Schedule 17 – Vitamins and minerals [Internet]. 2021 [cited 2024 Feb 25]. Available from: https://www.legislation.gov.au/F2015L00449/latest/text.

17. Dunlop E, James AP, Cunningham J, Rangan A, Daly A, Kiely M, et al. Vitamin D fortification of milk would increase vitamin D intakes in the australian population, but a more comprehensive strategy is required. Foods. 2022;11(9):1369.

18. Cashman KD, Kiely M. Tackling inadequate vitamin D intakes within the population: fortification of dairy products with vitamin D may not be enough. Endocrine. 2016;51(1):38-46.

19. Australian Bureau of Statistics. Australian Health Survey: users' guide, 2011-13. Canberra: Australian Bureau of Statistics; 2013 Jun. Report No.: 4363.0.55.001.

20. Australian Bureau of Statistics. Australian Aboriginal and Torres Strait Islander Health Survey: users' guide, 2012-13. Canberra: Australian Bureau of Statistics; 2013 Nov. Report No.: 4727.0.55.002.

21. Australian Bureau of Statistics. Australian Health Survey food model booklet [Internet]. Canberra: Australian Bureau of Statistics; 2012 [cited 2024 Feb 25]. Available from: https://www.ausstats.abs.gov.au/ausstats/subscriber.nsf/0/FF7E03F75EF8F5D9CA257E8400149FEA/$File/natsinpas%20food%20model%20booklet.pdf.

22. Australian Bureau of Statistics. Australian Health Survey bush tucker prompt card [Internet]. Canberra: Australian Bureau of Statistics; 2015 [cited 2023 Feb 25]. Available from: https://www.ausstats.abs.gov.au/ausstats/subscriber.nsf/0/6B5E429F3338729CCA257E840014A05C/$File/natsinpas%20prompt%20card%203%20bush%20tucker.pdf.

23. Food Standards Australia New Zealand. AUSNUT 2011-2013 [Internet]. 2016 [cited 2023 Feb 25]. Available from: https://www.foodstandards.gov.au/science-data/monitoringnutrients/ausnut/ausnutdatafiles#food-dietary.

24. Dunlop E, James AP, Cunningham J, Strobel N, Lucas RM, Kiely M, et al. Vitamin D composition of Australian foods. Food Chem. 2021;358:129836.

25. Dunlop E, Shepherd CCJ, Cunningham J, Strobel N, Lucas RM, Black LJ. Vitamin D composition of Australian game products. Food Chem. 2022;387:132965.

26. Food Standards Australia New Zealand. 25th Australian total diet study [Internet]. 2019 [cited 2023 Dec 05]. Available from: https://www.foodstandards.gov.au/publications/Documents/25thAustralianTotalDietStudy.pdf.

27. Australian Bureau of Statistics. Australian Aboriginal and Torres Strait Islander Health Survey: Nutrition Results - Food and Nutrients, 2012-13. Canberra: Australian Bureau of Statistics; 2015 Mar. Report No.: 4727.0.55.005.

28. Food Standards Australia New Zealand. Food Standards Code [Internet]. n.d. [cited 2023 Feb 25]. Available from: https://www.foodstandards.gov.au/code/Pages/default.aspx.

29. National Health and Medical Research Council. Australian Dietary Guidelines Summary. Canberra: National Health and Medical Research Council; 2013.

30. Food Standards Australia New Zealand. AUSNUT 2011-13 data files [Internet]. n.d. [cited 2024 Feb 25]. Available from: https://www.foodstandards.gov.au/science-data/monitoringnutrients/ausnut/ausnutdatafiles.

31. Machado PP, Steele EM, Levy RB, Sui Z, Rangan A, Woods J, et al. Ultra-processed foods and recommended intake levels of nutrients linked to non-communicable diseases in Australia: evidence from a nationally representative cross-sectional study. BMJ Open. 2019;9(8):e029544.

32. StataCorp. Stata statistical software. Version 17 [software]. 2021 [cited 2023 May 16]. Available from: https://www.stata.com/company/.

33. National Health and Medical Research Council, Australian Government Department of Health and Ageing, New Zealand Ministry of Health. Nutrient Reference Values for Australia and New Zealand including Recommended Dietary Intakes [Internet]. Canberra: National Health and Medical Research Council; 2006 [cited 2023 Feb 25]. Available from: www.nrv.gov.au.

34. Tebben PJ, Singh RJ, Kumar R. Vitamin D-mediated hypercalcemia: mechanisms, diagnosis, and treatment. Endocr Rev. 2016;37(5):521-47.

35. Dunlop E, Kiely ME, James AP, Singh T, Pham NM, Black LJ. Vitamin D food fortification and biofortification increases serum 25-hydroxyvitamin D concentrations in adults and children: an updated and extended systematic review and meta-analysis of randomized controlled trials. J Nutr. 2021;151(9):2622-35.

36. Neale RE, Wilson LF, Black LJ, Waterhouse M, Lucas RM, Gordon LG. Hospitalisations for falls and hip fractures attributable to vitamin D deficiency in older Australians. Br J Nutr. 2021;126(11):1682-6.

37. Australian Institute of Health and Welfare. Falls [Internet]. Canberra: Australian Institute of Health and Welfare; 2023 [cited 2024 Feb 25]. Available from: https://www.aihw.gov.au/reports/injury/falls.

38. Hiligsmann M, Burlet N, Fardellone P, Al-Daghri N, Reginster JY. Public health impact and economic evaluation of vitamin D-fortified dairy products for fracture prevention in France. Osteoporos Int. 2017;28(3):833-40.

39. Sandmann A, Amling M, Barvencik F, König HH, Bleibler F. Economic evaluation of vitamin D and calcium food fortification for fracture prevention in Germany. Public Health Nutr. 2017;20(10):1874-83.

40. Pasco JA, Wark JD, Carlin JB, Ponsonby AL, Vuillermin PJ, Morley R. Maternal vitamin D in pregnancy may influence not only offspring bone mass but other aspects of musculoskeletal health and adiposity. Med Hypotheses. 2008;71(2):266-9.

41. Petrenya N, Lamberg-Allardt C, Melhus M, Broderstad AR, Brustad M. Vitamin D status in a multi-ethnic population of northern Norway: the SAMINOR 2 Clinical Survey. Public Health Nutr. 2020;23(7):1186-200.

42. Wilson A, Wilson R, Delbridge R, Tonkin E, Palermo C, Coveney J, et al. Resetting the narrative in Australian Aboriginal and Torres Strait Islander nutrition research. Curr Dev Nutr. 2020;4(5):nzaa080.

43. Christidis R, Lock M, Walker T, Egan M, Browne J. Concerns and priorities of Aboriginal and Torres Strait Islander peoples regarding food and nutrition: a systematic review of qualitative evidence. Int J Equity Health. 2021;20:220.

44. Australian Bureau of Statistics. Under-reporting in the National Aboriginal and Torres Strait Islander nutrition survey. Canberra: Australian Bureau of Statistics; 2015 Mar. Report No.: 4727.0.55.002.

45. Macdiarmid J, Blundell J. Assessing dietary intake: who, what and why of under-reporting. Nutr Res Rev. 1998;11(2):231-53.

46. Malacova E, Cheang PR, Dunlop E, Sherriff JL, Lucas RM, Daly RM, et al. Prevalence and predictors of vitamin D deficiency in a nationally representative sample of adults participating in the 2011-2013 Australian Health Survey. Br J Nutr. 2019;121(8):894-904.

**Table 1.** Food fortification scenarios with vitamin D concentration for each food

| Scenario | Foods | Vitamin D concentration (µg/100 g) |
|---|---|---|
| 1 | *Scenario 1 includes foods and maximum vitamin D concentrations that are permitted for vitamin D fortification in Australia.* | |
| | Fluid milk (cow's milk, sheep's milk, goat's milk) and alternatives derived from legumes | 0.80 |
| | Butter, edible oil spreads and margarine | 16.00 |
| | Dried milk | 1.50 |
| | Cheese and cheese products and alternatives derived from legumes | 6.40 |
| | Yoghurts (with or without other foods) and alternatives derived from legumes | 1.07 |
| | Dairy desserts | 1.07 |
| | Beverages containing no less than 0.3% m/m protein derived from cereals, nuts, seeds, or a combination of those ingredients | 0.80 |
| | Formulated beverages (e.g., bottled water with added sugar, vitamins and minerals) | 0.42 |
| | NPSC-compliant RTEBCs | 8.33 |
| 2 | *Scenarios 2a-c include concentrations higher than those permitted in Australia. Scenario 2c includes bread, which is not permitted for fortification in Australia.* | |
| (a) | Fluid milk (cow's milk, sheep's milk, goat's milk) and alternatives derived from legumes | 1.00 |
| | Butter, edible oil spreads and margarine | 20.00 |
| | Dried milk | 1.50 |
| | Cheese and cheese products and alternatives derived from legumes | 6.40 |
| | Yoghurts (with or without other foods) and alternatives derived from legumes | 1.07 |
| | Dairy desserts | 1.07 |
| | Beverages containing no less than 0.3% m/m protein derived from cereals, nuts, seeds, or a combination of those ingredients | 1.00 |
| | Formulated beverages | 0.42 |
| (b) | Foods in Scenario 2a with the addition of: | |
| | NPSC-compliant RTEBCs | 8.33 |
| (c) | Foods in Scenario 2b with the addition of: | |
| | Bread (plain and sweet bread, rolls and buns, damper, bagels, focaccia, english muffins, and flatbread) | 1.70 |

NPSC, Nutrient Profiling Scoring Criterion; RTEBC, Ready-To-Eat Breakfast Cereal

**Table 2.** Vitamin D intakes of baseline and food fortification scenarios among Aboriginal and Torres Strait Islander peoples for ages ≥ 2 years[1]

| Age group (years) | Baseline | Scenario 1[2] | Scenario 2a[2] | Scenario 2b[2] | Scenario 2c[2] |
|---|---|---|---|---|---|
| | | Median (25th, 75th percentile) | | | |
| **Total** | | | | | |
| All ages | 2.0 (1.1, 3.6) | 5.6 (3.1, 9.0) | 5.2 (2.9, 9.0) | 6.2 (3.4, 10.0) | 7.7 (4.4, 11.9) |
| 2 - 3 | 1.2 (0.8, 2.1) | 7.6 (4.0, 10.6) | 6.3 (3.3, 11.9) | 8.3 (4.4, 11.9) | 9.6 (5.5, 13.7) |
| 4 - 8 | 1.8 (1.1, 3.0) | 5.7 (3.4, 8.6) | 4.9 (2.9, 8.2) | 6.5 (3.8, 9.3) | 8.0 (5.1, 11.2) |
| 9 - 13 | 2.3 (1.3, 4.0) | 6.3 (3.5, 9.0) | 5.6 (3.5, 9.4) | 7.1 (3.8, 10.3) | 8.5 (5.2, 12.4) |
| 14 - 18 | 1.9 (0.9, 3.8) | 4.9 (2.3, 9.3) | 5.2 (2.3, 8.2) | 5.8 (2.5, 10.3) | 6.8 (3.2, 12.2) |
| 19 - 30 | 2.2 (1.2, 4.0) | 5.2 (2.7, 9.0) | 5.4 (2.8, 9.4) | 5.9 (3.0, 9.9) | 7.1 (4.0, 11.7) |
| 31 - 50 | 2.2 (1.3, 3.8) | 5.1 (2.7, 8.8) | 5.1 (2.8, 8.9) | 5.8 (3.1, 9.4) | 7.4 (4.0, 11.0) |
| 51 - 70 | 2.0 (1.1, 3.4) | 5.8 (3.3, 9.0) | 5.4 (3.1, 9.0) | 6.4 (3.6, 10.0) | 8.1 (4.7, 11.6) |
| ≥ 71 | 1.0 (0.8, 2.4) | 4.3 (2.8, 7.5) | 3.7 (2.4, 6.0) | 4.6 (3.1, 8.0) | 6.0 (4.1, 9.3) |
| **Male** | | | | | |
| All ages | 2.2 (1.3, 4.1) | 6.3 (3.4, 10.1) | 5.9 (3.3, 9.6) | 6.9 (3.7, 11.1) | 8.4 (4.9, 12.9) |
| 2 - 3 | 1.3 (1.0, 1.8) | 7.6 (4.4, 10.7) | 6.5 (3.4, 12.7) | 8.5 (4.9, 12.0) | 10.2 (5.9, 13.7) |
| 4 - 8 | 1.8 (1.1, 3.6) | 5.7 (3.3, 9.5) | 4.7 (2.8, 8.4) | 6.0 (3.6, 10.3) | 8.2 (4.8, 11.4) |
| 9 - 13 | 2.5 (1.7, 4.2) | 7.3 (3.9, 10.1) | 6.5 (3.7, 10.1) | 8.2 (4.2, 11.2) | 9.8 (5.5, 13.1) |
| 14 - 18 | 2.4 (1.2, 4.0) | 6.8 (3.6, 11.0) | 6.3 (3.6, 9.9) | 7.7 (4.0, 11.8) | 9.6 (5.5, 13.6) |
| 19 - 30 | 2.4 (1.4, 4.8) | 5.9 (3.4, 10.1) | 6.2 (3.4, 10.7) | 6.4 (3.7, 11.5) | 8.2 (4.4, 13.8) |
| 31 - 50 | 2.5 (1.5, 4.4) | 5.8 (3.1, 9.7) | 5.8 (3.4, 9.4) | 6.5 (3.5, 10.5) | 8.1 (4.7, 13.1) |
| 51 - 70 | 1.9 (1.1, 3.5) | 6.1 (3.2, 9.0) | 5.1 (3.1, 9.1) | 6.4 (3.5, 10.0) | 8.1 (4.4, 12.1) |
| ≥ 71 | 1.6 (0.9, 4.0) | 6.5 (4.3, 11.1) | 5.7 (3.1, 9.3) | 7.4 (4.5, 12.8) | 8.5 (5.8, 13.7) |
| **Female** | | | | | |
| All ages | 1.9 (1.0, 3.2) | 5.1 (2.7, 8.0) | 4.9 (2.6, 8.1) | 5.6 (3.0, 8.9) | 7.0 (4.1, 10.4) |
| 2 - 3 | 1.1 (0.6, 2.4) | 7.3 (3.8, 10.1) | 5.9 (2.8, 11.9) | 8.1 (4.0, 11.9) | 9.3 (4.5, 13.7) |
| 4 - 8 | 1.8 (1.0, 2.8) | 5.9 (3.6, 8.2) | 5.6 (3.2, 8.2) | 6.6 (3.9, 8.8) | 7.9 (5.6, 10.5) |
| 9 - 13 | 2.2 (1.2, 3.4) | 5.5 (3.0, 8.3) | 5.2 (2.7, 8.2) | 6.0 (3.5, 9.4) | 8.2 (5.0, 11.1) |
| 14 - 18 | 1.4 (0.7, 2.9) | 3.8 (1.7, 6.2) | 4.3 (1.9, 6.6) | 4.3 (1.9, 6.9) | 5.3 (2.3, 8.1) |
| 19 - 30 | 2.0 (1.0, 3.4) | 4.3 (2.3, 7.7) | 4.4 (2.5, 8.2) | 4.7 (2.5, 8.4) | 6.2 (3.5, 9.9) |
| 31 - 50 | 1.9 (1.0, 3.3) | 4.6 (2.6, 7.5) | 4.7 (2.6, 7.8) | 5.2 (2.8, 8.5) | 6.6 (3.7, 9.9) |
| 51 - 70 | 2.0 (1.1, 3.4) | 5.7 (3.6, 9.3) | 5.9 (3.2, 8.9) | 6.4 (3.9, 10.3) | 8.1 (4.8, 11.5) |
| ≥ 71 | 1.0 (0.8, 2.3) | 2.9 (2.8, 7.2) | 3.1 (2.2, 4.4) | 3.3 (3.1, 7.7) | 4.2 (4.1, 7.7) |

[1] Weighted to the 2012-2013 Aboriginal and Torres Strait Islander population
[2] The modelling for Scenario 1 included foods and maximum vitamin D concentrations permitted for fortification in Australia: i) dairy products and alternatives, ii) butter/margarine/oil spreads, iii) formulated beverages, iv) selected ready-to-eat breakfast cereals. The modelling for Scenarios 2a-c included some vitamin D concentrations higher than permitted in Australia; Scenario 2c included bread, which is not permitted for vitamin D fortification in Australia. Scenario 2a: i) dairy products and alternatives, ii) butter/margarine/oil spreads, iii) formulated beverages. Scenario 2b: as per Scenario 2a plus selected ready-to-eat breakfast cereals. Scenario 2c: as per Scenario 2b plus bread.